\documentclass[aps,prxintelligence,reprint,nofootinbib,superscriptaddress,floatfix]{revtex4-2}

\usepackage{amsmath,amssymb}
\usepackage{graphicx}
\usepackage{booktabs}
\usepackage{siunitx}
\usepackage{xcolor}
\usepackage{hyperref}

\newcommand{\particlevit}{\textsc{ParticleViT}}
\newcommand{\omnilearned}{\textsc{OmniLearned}}
\newcommand{\rejfifty}{R_{50}}
\newcommand{\rejthirty}{R_{30}}

\begin{document}

\title{Predict before you train: Scaling Laws for particle physics foundation models}

\author{Jan-Lucas Uslu}
\affiliation{Department of Applied Physics, Stanford University, Stanford, CA 94305, USA}
\author{Benjamin Nachman}
\affiliation{Fundamental Physics Directorate, SLAC National Accelerator Laboratory, Menlo Park, CA 94025, USA}
\affiliation{Department of Physics, Stanford University, Stanford, CA 94305, USA}
\author{Christopher R\'e}
\affiliation{Department of Computer Science, Stanford University, Stanford, CA 94305, USA}

\date{\today}

\begin{abstract}
      The largest machine learning models in particle physics are also the most
      expensive to train, yet the return on scaling a given architecture cannot be
      estimated before that compute is spent. Scaling laws have been fit for jets,
      but none has yet been shown to predict the performance of models it was not
      fit on. We show that, for a generic transformer pretrained on collider jets,
      it can be forecast. Fitting a joint model-and-data scaling law on small models alone,
      spanning three orders of magnitude of training compute, we predict the loss of
      models trained afterward with more than one hundred times more compute to
      within one percent. We then connect the forecast to downstream physics
      performance: across two standard tagging benchmarks, lower pretraining loss
      yields systematically lower fine-tuning loss and higher background rejection
      after fine-tuning. Within this model family and these tasks, a compute
      budget can therefore be translated into expected physics performance
      before any large model is trained. The final frontier model is consistent with the
      published numbers for current state-of-the-art physics-aware foundation
      models trained on the same corpus, on accuracy, AUC, and quark/gluon rejection,
      with a residual edge for the physics-aware model only in the high-purity tail of
      top tagging. We release
      five pretrained models spanning multiple sizes, together with the complete
      training recipe and code.
\end{abstract}

\maketitle

\section{Introduction}
\label{sec:intro}

A recurring lesson of the past decade of machine learning is that simple
architectures, trained on large datasets with more compute, tend to outperform
carefully engineered alternatives~\cite{sutton2019bitter}. These gains turned
out to be predictable: the loss of a neural network falls as a power law in model
size, dataset size, and compute, and the fitted relationship extrapolates
reliably enough to plan training runs before launching
them~\cite{kaplan2020scaling,henighan2020scaling}. Hoffmann et~al., commonly
referred to as the Chinchilla study, showed how to allocate a fixed compute
budget between model size and training data by fitting a joint model-and-data
scaling law~\cite{hoffmann2022training}. Such laws are used to design and
budget large training runs.

Particle physics is an unusually favorable setting for this program.
Collider experiments and their simulations produce effectively unlimited
labeled and unlabeled data with a known generative process, and a growing body
of work is preparing both simulated and archival experimental data for machine
learning~\cite{cheng2026aiready}. The field has
begun building foundation models that learn
transferable representations from simulated
collisions~\cite{bhimji2025omnilearned,qu2022particle,birk2024omnijet,golling2024masked}.
Scaling laws have been fit for jets as
well~\cite{batson2025scaling,vigl2026neural,amram2026neural},
but always retrospectively: a law is fit to the full set of available runs and
its exponents are reported. Two properties that made scaling laws useful
elsewhere have not been shown for jets. No study has
fit a law on small models, committed to a prediction, and then trained the
large models to test it; nor has any connected a pretraining-loss law to the
downstream physics metrics that justify the compute. Without both, the return
on scaling a given architecture cannot be estimated in
advance, which matters in a field where the largest models are also the most
expensive to train.

We establish both here. We study \particlevit{}, a generic
transformer applied to the constituents of a jet with essentially no
physics-specific structure, pretrained on the roughly one billion jet
\omnilearned{} corpus~\cite{bhimji2025omnilearned}. Our contributions are the
following.

\begin{itemize}
      \item \textbf{A scaling law that extrapolates.} Fitting the joint
            model-and-data law introduced in the Chinchilla study only on models
            trained below $10^{19}$ FLOPs, we predict the loss
            of models trained afterward at more than one hundred times more
            compute, and the measured loss lands within about one percent of the
            prediction (Sec.~\ref{sec:extrapolation}). To
            our knowledge this is the first validated forecast, rather than
            retrospective fit, of a scaling law in particle physics.
      \item \textbf{Loss as a planning instrument.} Across two standard tagging
            tasks, lower pretraining loss yields systematically lower downstream
            fine-tuning loss and higher background rejection after fine-tuning
            (Sec.~\ref{sec:transfer}). Together with the first contribution this
            closes the chain from compute budget to predicted loss to expected
            performance on these tasks, before any large model is trained.
      \item \textbf{Open models.} We release five pretrained \particlevit{}
            models spanning the ladder (\particlevit{}-S through \particlevit{}-XL;
            Table~\ref{tab:checkpoints}), together with the full training recipe, code and
            fine-tuning protocol, so that the law and the models can be used
            directly (Sec.~\ref{sec:availability}).
\end{itemize}

Following the law to its frontier, we further benchmark the planned model
against \omnilearned{}, a state-of-the-art physics-aware foundation model
trained on the same corpus (Sec.~\ref{sec:sota}). Compared against the published
\omnilearned{} numbers, the two are consistent on accuracy and AUC and on
quark/gluon rejection, with a residual edge for the physics-aware model only in
the high-purity tail of top tagging. A by-product
of the scaling fit is the compute-optimal model size for the one billion jet
corpus, far smaller than the rule of thumb inherited from language models
(Sec.~\ref{sec:compute-optimal}). We argue in Sec.~\ref{sec:discussion} that
this reflects the low information content of a particle token,
which also reconciles our results with the saturation reported by prior jet
scaling studies~\cite{vigl2026neural,amram2026neural}.

For readers outside collider physics: top tagging asks whether a jet originates
from a boosted top
quark or from light-quark and gluon QCD radiation; quark/gluon asks whether a
jet was initiated by a quark or a gluon. We report background rejection
$\rejfifty = 1/\varepsilon_B$ at signal efficiency $\varepsilon_S = 0.5$, and
$\rejthirty$ at $\varepsilon_S = 0.3$, the standard figures of merit on these
benchmarks~\cite{kasieczka2019landscape,komiske2019energyflow}.

\section{Setup}
\label{sec:setup}

\subsection{Model: a low-inductive-bias transformer}

\particlevit{} treats each particle in a jet as a token and applies a standard
transformer to the resulting unordered set. The backbone follows
the design choices popularized by recent general-purpose language
models~\cite{olmo2025olmo3}: reordered normalization kept
outside the residual stream, RMSNorm in place of LayerNorm~\cite{zhang2019rmsnorm},
query-key normalization for attention
stability, as used at scale in~\cite{dehghani2023scaling}, and gated-linear-unit feedforward layers
with the $8/3$ width convention~\cite{shazeer2020glu}. We hold the
width-to-depth aspect ratio approximately constant near one hundred as we scale,
the value found optimal for transformers by Kaplan et~al.~\cite{kaplan2020scaling},
so that model size is varied along a single well-defined family. Readout is from
a single prepended class token, and there is no positional encoding, reflecting
that a jet's constituents form a set rather than a sequence. The architecture is
described in full in Sec.~\ref{sec:methods-arch}; the point for what follows is
that it deliberately omits the inductive biases (Lorentz equivariance, pairwise
interaction features, infrared and collinear safety) that physics-aware networks
build in~\cite{gong2022lorentznet,bogatskiy2022pelican,spinner2024lorentz}.

\subsection{Pretraining data and objective}

We pretrain on the \omnilearned{} dataset~\cite{bhimji2025omnilearned}, a collection of
seven jet datasets totaling roughly $1.06\times 10^{9}$ training jets; the
per-jet features and full layout are given in Sec.~\ref{sec:methods-data}. The
pretraining objective is classification over the flat label space defined by
the corpus: 200 jet process and flavor classes plus ten dataset-level classes.
These are jet-level categories, not a vocabulary of constituent particles.
We report the Gaussian-smoothed train loss as the
scaling quantity; sweep runs make less than one pass over the corpus, so this
loss is evaluated on examples the model has not seen before. The precise
definition is given in Sec.~\ref{sec:methods-loss}.

\subsection{Downstream tasks and baseline}

For downstream evaluation we use two widely used community benchmarks that are disjoint from the pretraining mix: the Top Quark Tagging Reference
Dataset~\cite{kasieczka2019landscape} and the Pythia8 quark/gluon
dataset~\cite{komiske2019energyflow}. We fine-tune the pretrained trunk with a
fresh binary head and report accuracy, AUC, and background rejection at
$\varepsilon_S = 0.5$ and $0.3$.

As a physics-aware baseline model we use \omnilearned{}~\cite{bhimji2025omnilearned},
an established foundation model that incorporates physics inductive bias and
reports state-of-the-art performance on these benchmarks. It is pretrained on
the same corpus, so the comparison holds the pretraining data fixed. We compare against the
metrics reported in the \omnilearned{} study throughout. To place both models
in the context of the wider field, we also reproduce the published tagger
results compiled in the \omnilearned{} study~\cite{bhimji2025omnilearned}.

\section{A predictive scaling law}
\label{sec:scaling}

\subsection{Loss definition and compute accounting}

We fit the scaling law on a geometric ladder of \particlevit{} models trained on
an IsoFLOP grid, varying both the number of trainable parameters $N$ and the
number of training jets $D$ at fixed compute. Here $N$ is the total trainable
parameter count (Sec.~\ref{sec:methods-loss}).
Training compute is accounted from the model size and the measured mean token
occupancy of the jets; the precise FLOP convention and per-jet token count are
given in Sec.~\ref{sec:methods-loss}.

\subsection{Functional form and fit}

\begin{figure*}[t]
      \centering
      \includegraphics[width=\textwidth]{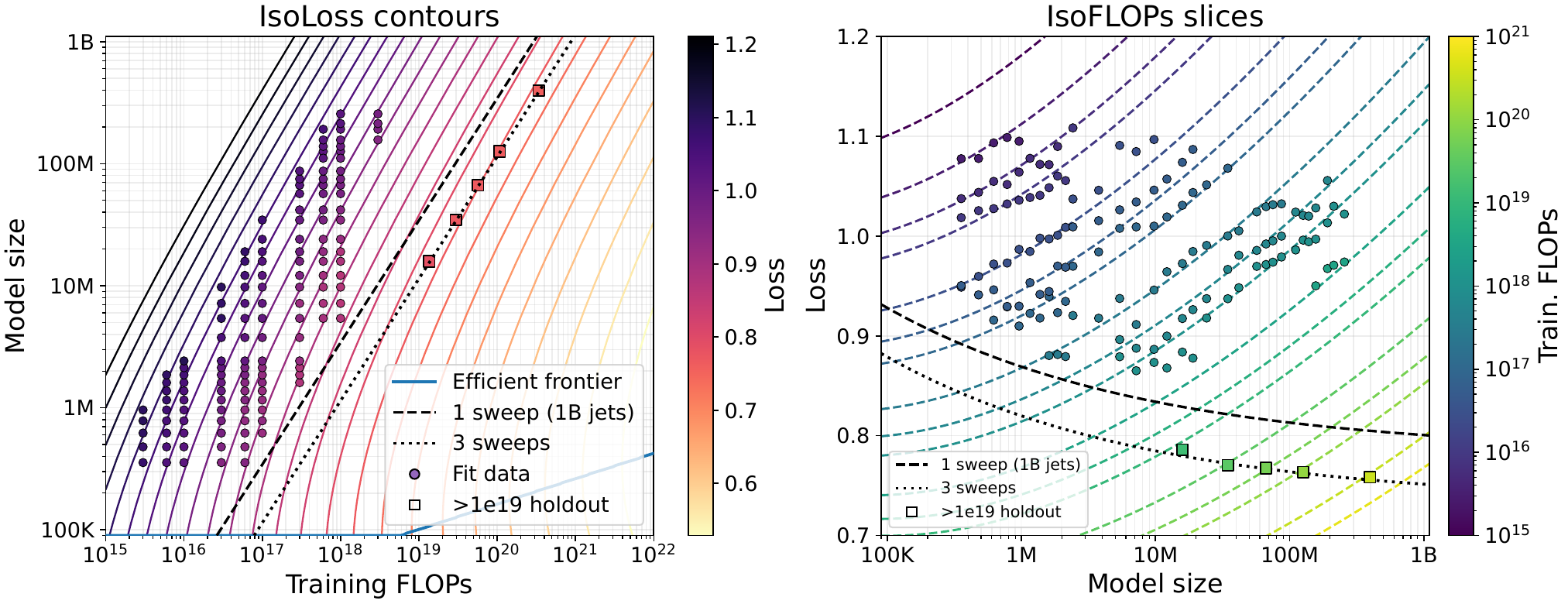}
      \caption{\textbf{A predictive scaling law.} IsoLoss contours in the
            (model size, training FLOPs) plane (left) and IsoFLOP slices of loss versus
            model size (right) for \particlevit{} on the \omnilearned{} corpus. Circles
            are the runs used in the fit; squares are the five held-out runs
            (Table~\ref{tab:extrap}), which lie on the contours and slices fit to the
            circles alone, showing that the law extrapolates. The lines labeled one and
            three sweeps mark, as
            a function of model size, the compute consumed by the corresponding number of
            passes over the full one billion jet corpus.}
      \label{fig:iso}
\end{figure*}

We model the pretraining loss with the Chinchilla parametric form
\begin{equation}
      \label{eq:scaling}
      L(N, D) \;=\; \frac{A}{N^{\alpha}} \;+\; \frac{B}{D^{\beta}} \;+\; L_\infty,
\end{equation}
where $L_\infty$ is the irreducible loss, and we estimate
$\{A, B, \alpha, \beta, L_\infty\}$ from the ladder. Figure~\ref{fig:iso}
visualizes the result as IsoLoss contours in the (model size, training FLOPs)
plane together with IsoFLOP slices of loss versus model size. We estimate the
parameters following Hoffmann et~al.~\cite{hoffmann2022training}, minimizing the
Huber loss ($\delta = 10^{-3}$) of the residual between predicted and measured
$\log L$ over the ladder with L-BFGS-B from a grid of initializations and
retaining the best optimum; the fit
gives $\alpha \approx 0.23$, $\beta \approx 0.06$, and $L_\infty$ unresolved
(Table~\ref{tab:fitparams}). The amplitude $A$ and the
exponent $\alpha$ are strongly degenerate, so we quote $\alpha$ alone. The fit
uses the same functional form as the Chinchilla study, whose language-model fit
gave $\alpha = 0.34$ and $\beta = 0.28$~\cite{hoffmann2022training}. The most
striking difference is therefore the much smaller data exponent in our fit.
This comparison is qualitative because the objectives, token definitions, and
data modalities differ. The fit
favors a negligible irreducible loss, $L_\infty$ rails to the lower bound of
the search ($10^{-12}$), but our data do not actually constrain it:
$L_\infty$ trades off against the data exponent $\beta$, and refitting the same
form under least-squares and robust (soft-$L_1$) objectives, or on $80\%$
subsamples, places $L_\infty$ anywhere from $0$ to $\approx 0.2$ with negligible
change in fit quality. We therefore report it as an upper limit,
$L_\infty \lesssim 0.2$ (consistent with zero), rather than a measured value. The
$210$-class objective has a strictly positive Bayes floor, but our data do not
resolve it, so we carry the fit's near-zero $L_\infty$ through to the
extrapolation and treat the probed regime as power-law, without reading the
bound as the true irreducible loss. The data exponent $\beta$ is small and, with
$L_\infty$ pinned at its floor, tightly determined (10th - 90th-percentile width
below $\pm0.003$); allowing a nonzero $L_\infty$ along the degeneracy above
widens it only to $\beta \lesssim 0.09$. The least-constrained direction is the
amplitude $A$ together with its degenerate partner $\alpha$, whose interval spans
roughly $0.14$ to $0.29$. The compute-optimal allocation
discussed below therefore inherits its uncertainty from $\alpha$, not from
$\beta$.

\begin{table}[t]
      \centering
      \caption{Fitted parameters of the Chinchilla-style scaling law
            [Eq.~\eqref{eq:scaling}], estimated on the $136$ fit points, one per (compute budget, model size) cell,
      across ten compute budgets all below $10^{19}$~FLOPs. Central values are the nominal fit
      (Huber loss on log-residuals, L-BFGS-B); parenthetical intervals on
      $\alpha$ and $\beta$ are the 10th and 90th
      percentiles obtained by refitting on 100 random 80\% subsamples of the fit
      points, the procedure of Hoffmann et~al.~\cite{hoffmann2022training}. The
      amplitude $A$ and the exponent $\alpha$ are strongly degenerate, so $A$ is
      poorly constrained and the two are quoted as point estimates; the
      well-constrained combinations are the normalized amplitudes
      $A/N_0^{\alpha} = 0.057$ and $B/D_0^{\beta} = 0.93$ at pivots
      $N_0 = 6.6\times10^{6}$ and $D_0 = 6.0\times10^{7}$. The irreducible loss
      $L_\infty$ is unresolved: the fit rails it to the search floor ($10^{-12}$),
      but it is degenerate with $\beta$ and, across least-squares and robust
      objectives, consistent with anything from $0$ to $\approx 0.2$, so we report
      it as an upper limit ($L_\infty \lesssim 0.2$) and carry the fit's near-zero
      value through to the extrapolation.}
      \label{tab:fitparams}
      \begin{ruledtabular}
            \begin{tabular}{l l l}
                                        & Symbol     & Value                    \\
                  \colrule
                  Parameter coefficient & $A$        & $1.95$                   \\
                  Parameter exponent    & $\alpha$   & $0.225~(0.14,\,0.29)$    \\
                  Data coefficient      & $B$        & $2.67$                   \\
                  Data exponent         & $\beta$    & $0.0591~(0.057,\,0.063)$ \\
                  Irreducible loss      & $L_\infty$ & $\lesssim 0.2$ (unresolved) \\
            \end{tabular}
      \end{ruledtabular}
\end{table}

\subsection{Compute-optimal allocation favors data}
\label{sec:compute-optimal}

The fit also gives the compute-optimal model size for the full corpus, which is
far smaller than the language-model expectation. The
language-model rule of thumb of roughly one parameter per twenty training
tokens~\cite{hoffmann2022training} would place the optimum near
$5\times 10^{7}$ parameters for a corpus of this token count; the fitted law
instead places it well below that. We interpret
the gap as reflecting information density: a particle token, drawn from a
handful of kinematic features and a small identification alphabet, carries less
information than a language token drawn from a large vocabulary, so less model
capacity is needed to absorb a given number of tokens.
The precise value of the optimum leans on the weakly constrained parameter
exponent $\alpha$, so we read it only as an order of magnitude: the fitted law
places the compute-optimal size at or below $10^{5}$ parameters at the
full-corpus budget, below the smallest model in our grid ($3.6\times 10^{5}$
parameters) and at the floor of our search, so the exact value is an
extrapolation that should not be overinterpreted. The more robust statement is
the compute-optimal exponent itself: with $N_{\mathrm{opt}}\propto C^{a}$, the
fit gives $a = \beta/(\alpha+\beta) = 0.21$, with a 10th - 90th-percentile range
of $(0.17,\,0.30)$ from the subsample bootstrap, far below the near-equal
$a\approx0.5$ of language models~\cite{hoffmann2022training}, so compute is best
spent overwhelmingly on data rather than on model size; correspondingly,
$D_{\mathrm{opt}}\propto C^{0.79}$ for the nominal fit. Either way the gap to
the language-model rule of thumb exceeds two orders of magnitude, large enough
that the qualitative conclusion, a compute-optimal model far smaller than the
language-model expectation, is robust even though the precise optimum is not.

The small $\beta$ has two distinct implications that should not be conflated.
At fixed model size, loss improves only slowly as the number of training jets
increases, so a meaningully lower loss requires a large increase in data and
compute. At fixed total compute, however, the fitted balance favors increasing
the number of training examples much faster than increasing model capacity.
Here $D$ counts processed jets. The sweep runs use less than one pass over the
corpus, while the held-out runs show that the forecast remains accurate through
roughly 2.8 passes; these experiments do not separately measure the value of a
new unique jet versus another pass over an existing one.
The fit alone cannot identify the mechanism: low signal-to-noise, redundant
examples, label ambiguity, and intrinsically stochastic radiation could all
produce weak data scaling. Our information-density interpretation is therefore
a hypothesis, not a measurement of signal-to-noise and left for future work to test.

\subsection{Held-out extrapolation}
\label{sec:extrapolation}

A scaling law is only useful if it predicts models it was not fit
on. We fit Eq.~\eqref{eq:scaling} using only models trained with fewer than
$10^{19}$ FLOPs (136 distinct (compute budget, model size) cells across ten compute
budgets, the largest at $3\times 10^{18}$ FLOPs, each cell swept over learning rate and
seed with the lowest-loss run kept; Sec.~\ref{sec:methods-data}),
then use the fit to predict the loss of five models
trained afterward at higher compute, up to $3.4\times 10^{20}$ FLOPs, more than
one hundred times beyond the largest budget in the fit set. The predicted and
measured losses agree closely (Table~\ref{tab:extrap}): the held-out models
land within one percent of their predicted loss, with a maximum relative error
of $0.66\%$ and a mean of $0.26\%$.

\begin{table}[t]
      \centering
      \caption{\textbf{The scaling law extrapolates.} Predicted versus measured
            pretraining loss for the five models trained after the fit, which used only
            runs at budgets up to $3\times10^{18}$~FLOPs. Predictions are from
            Eq.~\eqref{eq:scaling} with the parameters of Table~\ref{tab:fitparams},
            evaluated at each model's parameter count $N$ and training tokens. All five
            held-out models, trained at up to $113\times$ the largest compute in the fit
            set, are predicted to within one percent. These are the released
            \particlevit{}-S through \particlevit{}-XL checkpoints, shown as open squares
            in Fig.~\ref{fig:iso}.}
      \label{tab:extrap}
      \begin{ruledtabular}
            \begin{tabular}{r r c c r}
                  $N$    & $C$~[FLOPs]        & Measured & Predicted & Rel.\ err \\
                  \colrule
                  $16$M  & $1.4\times10^{19}$ & $0.7859$ & $0.7807$  & $-0.66\%$ \\
                  $35$M  & $3.0\times10^{19}$ & $0.7706$ & $0.7732$  & $+0.34\%$ \\
                  $67$M  & $5.7\times10^{19}$ & $0.7675$ & $0.7678$  & $+0.04\%$ \\
                  $126$M & $1.1\times10^{20}$ & $0.7634$ & $0.7633$  & $-0.02\%$ \\
                  $397$M & $3.4\times10^{20}$ & $0.7583$ & $0.7567$  & $-0.22\%$ \\
            \end{tabular}
      \end{ruledtabular}
\end{table}

\section{Pretraining loss tracks held-out benchmarks}
\label{sec:transfer}

\begin{figure*}[t]
      \centering
      \includegraphics[width=\textwidth]{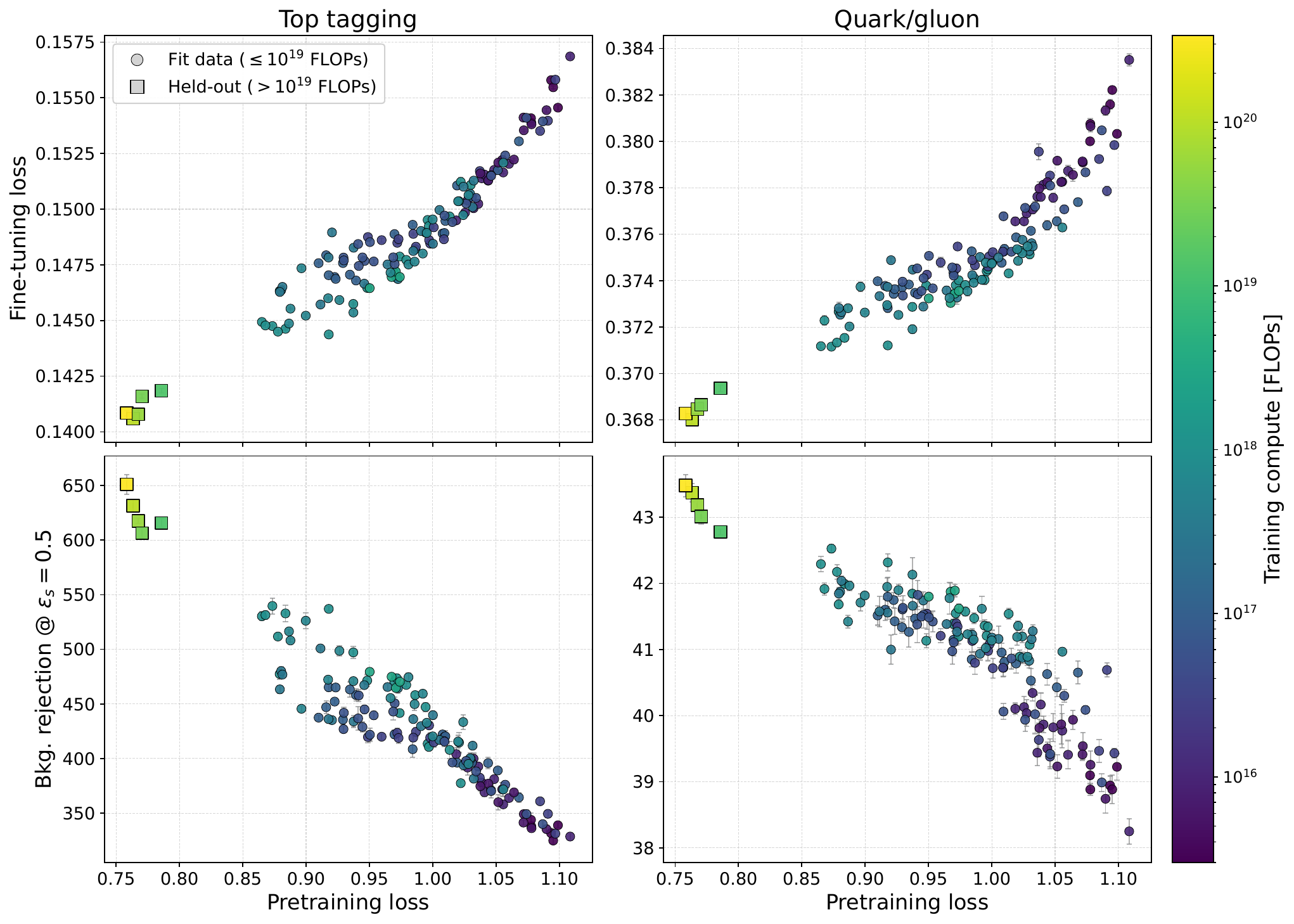}
      \caption{\textbf{Pretraining loss tracks downstream performance.} Downstream
            fine-tuning loss (top row) and background rejection at $\varepsilon_S = 0.5$
            (bottom row) versus pretraining loss, for top tagging (left) and quark/gluon
            (right), colored by training compute. Each point is one pretrained checkpoint,
            plotted as the mean over five independent fine-tuning runs
            (Sec.~\ref{sec:methods-data}); grey error bars give the standard deviation
            across those runs. Squares mark the five held-out frontier models (above
            $10^{19}$~FLOPs), the same models drawn as squares in Fig.~\ref{fig:iso};
            circles are the lower-compute models. Lower pretraining loss yields both lower
            downstream loss and higher rejection, ordered by compute.}
      \label{fig:transfer}
\end{figure*}

A scaling law for cross-entropy loss is useful to physicists only if the loss
tracks physics performance. We find that it does. The transfer is visible
first at the level of the loss itself: a checkpoint with lower pretraining loss
fine-tunes to a lower downstream task loss on both benchmarks, consistently and
ordered by training compute (Fig.~\ref{fig:transfer}, top row). It then carries
through to the physics metric: models with lower pretraining loss achieve higher
downstream background rejection after fine-tuning, with the same consistent,
compute-ordered trend (Fig.~\ref{fig:transfer}, bottom row), measured over
$141$ pretrained checkpoints per task (the $136$ fit-set models plus the five
held-out frontier models), where each checkpoint is summarized by
the mean over five independent fine-tuning runs. Within the model family,
compute range, and benchmarks studied here, two consequences follow. First, the chain compute
$\rightarrow$ pretraining loss (Sec.~\ref{sec:scaling}) $\rightarrow$
downstream rejection (this section) lets a physics-performance target be
translated into an estimated compute budget in advance. Second, because the pretraining
loss is the common currency, the inexpensive small-model runs used to fit the
scaling law already indicate the downstream performance expected on these
tasks. This is dataset and representation transfer to held-out samples, not
evidence that loss alone predicts performance on arbitrary new physics
questions.
The fine-tuning protocol is specified in Sec.~\ref{sec:methods-data}.

\section{Benchmarking the planned frontier model}
\label{sec:sota}

Having shown that the law forecasts the loss (Sec.~\ref{sec:extrapolation}) and
that the loss predicts physics performance (Sec.~\ref{sec:transfer}), we ask
how good the planned frontier model actually is. We compare \particlevit{}
against \omnilearned{} (Sec.~\ref{sec:setup}) on the same pretraining corpus, so
the comparison holds the data fixed while the two models differ in architecture
and in the physics inductive bias they encode.
Tables~\ref{tab:benchmarks_top} and~\ref{tab:benchmarks_qg} report accuracy,
AUC, and background rejection at $\varepsilon_S = 0.5$ and $0.3$ for the five
fully trained \particlevit{} models, \omnilearned{}, and the published taggers
compiled in Ref.~\cite{bhimji2025omnilearned}.

On the bulk classification metrics \particlevit{} is consistent with the published
\omnilearned{} numbers: at the frontier
\particlevit{} (397M) and \omnilearned{} (423M) reach similar accuracy
($0.9433$ against $0.944$) and AUC ($0.9875$ against $0.9880$), and the
agreement holds across the ladder. The physics-aware model retains a measurable
edge only in the high-purity tail of the top-tagging classifier, where rare
radiation patterns are scarcest: it leads in top-tagging rejection by about
five percent at the frontier ($\rejfifty = 688 \pm 9$ against $651 \pm 9$), and
the gap grows deeper into the tail, from $\rejfifty$ to $\rejthirty$. On
quark/gluon even this residual edge closes: \particlevit{} is consistent with
the best reported \omnilearned{} rejection from $67$M onward
($\rejfifty = 43.2 \pm 0.1$ against $43.2 \pm 0.1$), and the larger models sit
at or above it within one to two standard deviations. A generic transformer
pretrained at scale therefore reaches performance consistent with the published
numbers for the physics-aware baseline on
these benchmarks, with a residual advantage for encoded physics confined to the
extreme top-tagging rejection tail.

Placed against the broader field, both
same-corpus models rank among the strongest published taggers on either
benchmark, although individual columns are led by specialized architectures
trained under different protocols: the fine-tuned ParT attains the
highest top-tagging $\rejfifty$ ($691\pm15$) and L-GATr the highest
top-tagging accuracy, while \particlevit{} holds the best quark/gluon
$\rejthirty$ ($112.2\pm0.9$).

\begin{table}[tbp]
      \centering
      \caption{Performance on the Top Quark Tagging Reference
            Dataset~\cite{kasieczka2019landscape}. $\rejfifty$ and $\rejthirty$ are
            background rejections $1/\varepsilon_B$ at $50\%$ and $30\%$ signal
            efficiency. Each reference tagger is cited at its original publication, and the
            quoted numbers are the values compiled in the \omnilearned{}
            study~\cite{bhimji2025omnilearned}; the suffix ``f.t.'' marks a model
            pretrained on a foundation corpus and then fine-tuned.
            \omnilearned{} is the physics-aware foundation model trained on the same
            corpus as \particlevit{}; the \particlevit{} rows are the five fully trained
            models of the extrapolation test (batch size 16384, roughly $2.8$ passes over
            the corpus), each the mean $\pm$ standard deviation over five independent
            fine-tuning runs and summarized by the best test value over the fixed
            training trajectory (Sec.~\ref{sec:methods-data}). All \particlevit{} rows are fine-tuned from the pretrained trunk with a fresh
            binary head; the quoted \particlevit{} uncertainties are the standard deviation
            over five independent fine-tuning runs. The reference and \omnilearned{}
            uncertainties are cited from their source publications and may follow
            different conventions. The best value in each column is shown
            in bold; entries tied at the displayed precision are all bold.}
      \label{tab:benchmarks_top}
      \resizebox{\columnwidth}{!}{%
      \begin{minipage}{1.15\columnwidth}
      \begin{ruledtabular}
            \begin{tabular}{l cccc}
                  Model                 & Acc             & AUC             & $\rejfifty$             & $\rejthirty$              \\
                  \colrule
                  ResNeXt-50~\cite{qu2020particlenet} & 0.936           & 0.9837          & $302\pm5$               & $1147\pm58$               \\
                  P-CNN~\cite{qu2020particlenet} & 0.930           & 0.9803          & $201\pm4$               & $759\pm24$                \\
                  PFN~\cite{komiske2019energyflow} & n/r             & 0.9819          & $247\pm3$               & $888\pm17$                \\
                  ParticleNet~\cite{qu2020particlenet} & 0.940           & 0.9858          & $397\pm7$               & $1615\pm93$               \\
                  JEDI-net~\cite{moreno2020jedinet} & 0.9300          & 0.9807          & n/r                     & $774.6$                   \\
                  PCT~\cite{mikuni2021pct} & 0.940           & 0.9855          & $392\pm11$              & $1559\pm98$               \\
                  LGN~\cite{bogatskiy2020lgn} & 0.929           & 0.964           & n/r                     & $435\pm95$                \\
                  rPCN~\cite{shimmin2021particle} & n/r             & 0.9845          & $364\pm9$               & $1642\pm93$               \\
                  LorentzNet~\cite{gong2022lorentznet} & 0.942           & 0.9868          & $498\pm18$              & $2195\pm173$              \\
                  PELICAN~\cite{bogatskiy2022pelican} & 0.9425          & 0.9869          & n/r                     & $2289\pm204$              \\
                  ParT~\cite{qu2022particle} & 0.940           & 0.9858          & $413\pm16$              & $1602\pm81$               \\
                  ParT-f.t.~\cite{qu2022particle} & 0.944           & 0.9877          & $\boldsymbol{691\pm15}$ & $2766\pm130$              \\
                  Mixer~\cite{hammad2024mixer} & n/r             & 0.9859          & $416\pm5$                   & n/r                       \\
                  MIParT~\cite{wu2024mipart} & 0.942           & 0.9868          & $505\pm8$               & $2010\pm97$               \\
                  MIParT-f.t.~\cite{wu2024mipart} & 0.944           & 0.9878          & $640\pm10$              & $2789\pm133$              \\
                  L-GATr~\cite{brehmer2024lgatr} & 0.9423          & 0.9870          & $540\pm20$              & $2240\pm70$               \\
                  L-GATr-f.t.~\cite{brehmer2024lgatr} & \textbf{0.9446} & 0.9879          & $651\pm11$              & $2894\pm84$               \\
                  PET~\cite{mikuni2025pet,mikuni2024omnilearn} & 0.938           & 0.9848          & $340\pm12$              & $1318\pm39$               \\
                  OmniLearn~\cite{mikuni2025pet,mikuni2024omnilearn} & 0.942           & 0.9872          & $568\pm9$               & $2647\pm192$              \\
                  PET v2-s~\cite{bhimji2025omnilearned} & 0.9427          & 0.987           & $505\pm14$              & $2167\pm153$              \\
                  PET v2-m~\cite{bhimji2025omnilearned} & 0.9423          & 0.987           & $482\pm11$              & $1861\pm61$               \\
                  \colrule
                  \omnilearned{} (3M)~\cite{bhimji2025omnilearned} & 0.944           & 0.9875          & $565\pm12$              & $2637\pm128$              \\
                  \omnilearned{} (58M)~\cite{bhimji2025omnilearned} & 0.944           & \textbf{0.9880} & $656\pm12$              & $3208\pm176$              \\
                  \omnilearned{} (423M)~\cite{bhimji2025omnilearned} & 0.944           & \textbf{0.9880} & $688\pm9$               & $\boldsymbol{3486\pm157}$ \\
                  \colrule
                  \particlevit{} (16M)  & 0.9431          & 0.9873          & $616\pm4$               & $2707\pm29$               \\
                  \particlevit{} (35M)  & 0.9430          & 0.9874          & $606\pm5$               & $2928\pm76$               \\
                  \particlevit{} (67M)  & 0.9434          & 0.9875          & $618\pm2$               & $2903\pm74$               \\
                  \particlevit{} (126M) & 0.9434          & 0.9876          & $631\pm6$               & $3042\pm47$               \\
                  \particlevit{} (397M) & 0.9433          & 0.9875          & $651\pm9$               & $3008\pm104$              \\
            \end{tabular}
      \end{ruledtabular}%
      \end{minipage}}
\end{table}

\begin{table}[tbp]
      \centering
      \caption{Performance on the Pythia8 quark/gluon
            dataset~\cite{komiske2019energyflow}. Columns, model families, and
            conventions follow Table~\ref{tab:benchmarks_top}; reference taggers are cited
            at their original publications, with numbers compiled in the \omnilearned{}
            study~\cite{bhimji2025omnilearned}. The best value in each column is shown in
            bold.}
      \label{tab:benchmarks_qg}
      \resizebox{\columnwidth}{!}{%
      \begin{minipage}{1.15\columnwidth}
      \begin{ruledtabular}
            \begin{tabular}{l cccc}
                  Model                 & Acc             & AUC             & $\rejfifty$               & $\rejthirty$              \\
                  \colrule
                  P-CNN~\cite{qu2020particlenet} & 0.827           & 0.9002          & $34.7$                    & $91.0$                    \\
                  PFN~\cite{komiske2019energyflow} & n/r             & 0.9005          & $34.7\pm0.4$              & n/r                       \\
                  ParticleNet~\cite{qu2020particlenet} & 0.840           & 0.9116          & $39.8\pm0.2$              & $98.6\pm1.3$              \\
                  rPCN~\cite{shimmin2021particle} & n/r             & 0.9081          & $38.6\pm0.5$              & n/r                       \\
                  ParT~\cite{qu2022particle} & 0.840           & 0.9121          & $41.3\pm0.3$              & $101.2\pm1.1$             \\
                  ParT-f.t.~\cite{qu2022particle} & 0.843           & 0.9151          & $42.4\pm0.2$              & $107.9\pm0.5$             \\
                  PET~\cite{mikuni2025pet,mikuni2024omnilearn} & 0.837           & 0.9110          & $39.92\pm0.1$             & $104.9\pm1.5$             \\
                  OmniLearn~\cite{mikuni2025pet,mikuni2024omnilearn} & 0.844           & 0.9159          & $\boldsymbol{43.7\pm0.3}$ & $107.7\pm1.5$             \\
                  PET v2-s~\cite{bhimji2025omnilearned} & 0.842           & 0.9137          & $41.7\pm0.4$              & $104.4\pm1.3$             \\
                  PET v2-m~\cite{bhimji2025omnilearned} & 0.841           & 0.9135          & $41.3\pm0.6$              & $103.8\pm1.1$             \\
                  \colrule
                  \omnilearned{} (3M)~\cite{bhimji2025omnilearned} & 0.844           & 0.9153          & $42.9\pm0.2$              & $108.0\pm0.5$             \\
                  \omnilearned{} (58M)~\cite{bhimji2025omnilearned} & \textbf{0.845}  & \textbf{0.9162} & $43.2\pm0.1$              & $111.2\pm1.5$             \\
                  \colrule
                  \particlevit{} (16M)  & 0.8437          & 0.9150          & $42.8\pm0.0$              & $110.3\pm0.4$             \\
                  \particlevit{} (35M)  & 0.8438          & 0.9154          & $43.0\pm0.1$              & $108.9\pm0.5$             \\
                  \particlevit{} (67M)  & 0.8441          & 0.9154          & $43.2\pm0.1$              & $109.8\pm0.6$             \\
                  \particlevit{} (126M) & 0.8443          & 0.9157          & $43.4\pm0.1$              & $\boldsymbol{112.2\pm0.9}$ \\
                  \particlevit{} (397M) & 0.8441          & 0.9155          & $43.5\pm0.2$              & $110.2\pm0.7$             \\
            \end{tabular}
      \end{ruledtabular}%
      \end{minipage}}
\end{table}

\section{Discussion}
\label{sec:discussion}

\paragraph{Plannable scaling for particle physics.}
Because the pretraining loss extrapolates and predicts downstream rejection, a
practitioner can fit the law on a modest sweep of small models and forecast both
the loss and the physics performance of a much larger model before committing
the compute to train it. For the model family and benchmarks tested here, a
physics target can be translated into an estimated compute request, and a
compute budget into an expected rejection. Applying this chain to a new task
requires verifying that pretraining loss predicts that task's metric; it does
not follow from the scaling law alone. The fit-then-forecast protocol is not
specific to jets: any data-rich scientific domain that can afford a sweep of
small models could test the same chain before budgeting a large run, which is
the broader methodological point of this work.

\paragraph{The information content of physical data.}
The compute-optimal model for one billion jets sits far below the
language-model expectation (Sec.~\ref{sec:compute-optimal}). We read this as a statement about the information
carried by a particle token relative to a language token. The fitted exponents
in Eq.~\eqref{eq:scaling} encode how quickly model capacity must grow with data,
and a lower information density per token shifts the compute-optimal frontier
toward smaller models. This connects the scaling behavior of a physical dataset
to its information content, a link of interest beyond collider physics. We
emphasize that this information-density reading is a hypothesis motivated by the
fit rather than a quantity our data measure; testing it would require comparing
scaling exponents across data modalities, which we leave to future work.

\paragraph{Scale versus recipe.}
A natural concern is whether the frontier result reflects scale or our training
recipe (BF16 precision, RMSNorm, variable-length attention, Yeo-Johnson input
normalization). We hold the recipe fixed across the entire ladder, so within
\particlevit{} scale is the only variable. Against \omnilearned{} the shared
quantity is the pretraining corpus; the models otherwise differ in architecture
and inductive bias, so we read the comparison as a contrast between two model
families on common data, not as an isolated measurement of inductive bias.
We have not fit or prospectively tested a scaling law for \omnilearned{}, so the
benchmark comparison neither explains scaling artifacts in that family nor
shows that every architecture should be scaled according to the
\particlevit{} allocation. Instead, it motivates applying the same
fit-then-forecast test separately to each model family before committing a
larger compute budget.

\paragraph{Relation to prior scaling studies.}
Several earlier studies report scaling laws for jets. Batson and Kahn
established that the test
loss of supervised top taggers follows power laws in training-set size, with
exponents that differ between classifiers, so that the best classifier can
change as the dataset grows~\cite{batson2025scaling}. Vigl and collaborators
derive compute-optimal scaling laws for supervised tagging on JetClass and
identify an asymptotic performance limit that compute consistently
approaches~\cite{vigl2026neural}, and Amram and collaborators find that the
loss of generative jet models saturates with data and compute faster than its
language-model counterpart~\cite{amram2026neural}. A complementary line of work
shows that the trade-off between model size and dataset size can itself be
engineered through the composition of the pretraining
corpus~\cite{uslu2026engineering}. None of these conclusions
contradicts ours, for three reasons. First, Eq.~\eqref{eq:scaling} carries an
irreducible term $L_\infty$ that represents exactly such a floor, and the
saturation those studies observe is the approach to it. Our fit does not
resolve $L_\infty$ over the probed range and is consistent with zero
(Sec.~\ref{sec:scaling}), so within our regime the loss is still in its
power-law descent; our claim is not that the loss falls forever but that this
descent is predictable, and that performance comparable to the published
physics-aware numbers is reached well before any floor is hit. Consistent with this, where the floor
itself has been computed, current taggers still sit measurably below the
information-theoretic optimum of the classification
task~\cite{geuskens2025fundamental}, so the irreducible loss is not yet the
limiting factor. Second, the object of study differs: prior laws are fit to
supervised performance on a single task, whereas we fit the pretraining loss
of a single model family and show separately that it transfers to downstream
metrics, which is the combination that enables planning. Third, the regimes
differ: the prior studies train on corpora roughly an order of magnitude
smaller than the billion-jet bundle used here, so an apparent saturation can
be data-limited rather than physics-limited, consistent with the finding that
richer inputs raise the saturation limit~\cite{vigl2026neural} and with the
data-repetition analysis of Ref.~\cite{muennighoff2023scaling}. The
information-density reading of Sec.~\ref{sec:compute-optimal} explains this:
tokens that carry little information are learned by
small models and approach the irreducible loss quickly, which is why
saturation is prominent on smaller corpora, and why on a larger corpus the
same approach remains in its predictable power-law regime.

\paragraph{Limitations.}
The study covers a single physics domain (jets) and two downstream benchmarks;
the scaling law is validated over a finite compute range and a single holdout
threshold; the downstream labels overlap semantically with the pretraining
taxonomy; the reported fine-tuning summaries select the best value along each
test trajectory rather than selecting a checkpoint on validation alone; and
several architectural choices are reported as design decisions rather than
fully ablated. Natural extensions include rerunning model selection on
validation only, pushing the holdout further, adding tasks and detectors,
growing the corpus, and training the
compute-optimal frontier model predicted by the fit.

\section{Methods}
\label{sec:methods}

\subsection{Loss and FLOP definitions}
\label{sec:methods-loss}

\paragraph{Training objective.} The pretraining objective is plain softmax
cross-entropy over the 210-class \omnilearned{} label space, computed from the
class-token logits, with no label smoothing. To keep the logits bounded we add
an output $z$-loss~\cite{chowdhery2023palm},
$\mathcal{L} = \mathcal{L}_{\mathrm{CE}} +
      \lambda_z\,\mathbb{E}\!\left[\log^2 Z\right]$ with
$Z = \sum_k e^{z_k}$ and $\lambda_z = 10^{-5}$. The $z$-loss term is excluded
from the scaling quantity; only the cross-entropy term enters the fits.

\paragraph{Smoothed loss.} At every optimizer step we record the cross-entropy
of the current training batch, averaged over all data-parallel ranks. Because
the IsoFLOP sweep is capped at $10^{9}$ jets per run, less than one pass over
the corpus, no training example is repeated within a sweep run and the
training loss is computed on examples the model has not seen before; the five
large held-out runs make roughly $2.8$ passes, and we report the same
smoothed training loss for them. The per-step loss curve is smoothed with a
Gaussian kernel of width $\sigma = 250$ steps, truncated to a window of
$1001$ steps and renormalized over valid entries, and the loss assigned to a
run is the value of the smoothed curve at the final training step. For each
(compute budget, model size) cell of the grid we keep the best such loss over
the learning-rate grid and repeated seeds.

\paragraph{Compute accounting.} We account training compute with the standard
approximation of six FLOPs per parameter per training
token~\cite{kaplan2020scaling,hoffmann2022training}. Each jet provides a fixed
sequence of $L = 150$ particle slots plus one class token, of which a fraction
$f = 0.31$ of the particle slots is occupied on average, so a jet contributes
$fL + 1 \approx 47.5$ tokens. A run's training compute is therefore
\begin{equation}
      C \;=\; 6\,N\,\bigl(fL + 1\bigr)\,B\,S
\end{equation}
for batch size $B$ and $S$ optimizer steps, and the number of steps for each
IsoFLOP run is set by inverting this relation.

\paragraph{Parameter convention.} The model size $N$ entering
Eq.~\eqref{eq:scaling} is the total trainable parameter count, evaluated in
closed form from the architecture: the split per-particle input stem, the
class token, the transformer blocks, and the 210-way classification head.
Unlike language models, \particlevit{} has no large vocabulary embedding: the
input stem contributes only $23\,d$ parameters at width $d$, well below one
percent of the total, so the distinction between total and non-embedding
parameters~\cite{kaplan2020scaling} is immaterial here.

\subsection{Architecture}
\label{sec:methods-arch}

\particlevit{} embeds each particle through a split input stem (continuous
kinematics, a categorical particle-identification embedding, and separate
vertexing features), prepends a single learnable class token, and applies a
stack of transformer blocks in the reordered-RMSNorm, query-key-norm, SwiGLU
design of Sec.~\ref{sec:setup}, with $d_{ff} = \tfrac{8}{3}\,d_{\mathrm{model}}$.
Weights use the truncated-normal initialization of recent open language
models~\cite{olmo2025olmo3}. There is no
positional encoding and no depth-scaled residual initialization; stability comes
from the reordered-norm and query-key-norm combination. The ladder spans depths
3 to 14 and widths chosen to keep the aspect ratio near one hundred.

\subsection{Numerical precision and training stability}
\label{sec:numerics}

We train and store features in BF16. An
output $z$-loss term keeps the logits bounded and stabilizes
training~\cite{chowdhery2023palm}. Variable-length attention with sequence packing removes all padding tokens by
concatenating the jets in a batch into a single sequence, which reduces training
time by about \SI{70}{\percent} with no loss in accuracy and is a substantial
part of what makes the scaling sweep affordable.

\subsection{Datasets and training details}
\label{sec:methods-data}

\paragraph{Pretraining corpus.} We pretrain on the \omnilearned{}
dataset~\cite{bhimji2025omnilearned}, the union of seven jet datasets totaling
approximately $1.06\times 10^{9}$ training jets, with about $10^{8}$
validation and $7\times 10^{7}$ test jets held out. Each jet is stored as up
to 150 particles with nine per-particle features: the kinematic set
$(\Delta\eta, \Delta\phi, \log p_T, \log E)$, a categorical
particle-identification code, and four vertexing features that are zeroed
where a subset does not provide them. Labels live in the flat 210-class space
defined by the corpus: 200 labels distinguish jet-producing processes and
flavors, including fine-grained JetClass2 categories, and ten identify
dataset-level categories. The two downstream benchmark files are not included
in any pretraining split, even though top/QCD and quark/gluon concepts occur in
the pretraining taxonomy. The data are converted once into globally shuffled
BF16 shards and streamed during training; a frozen Yeo-Johnson transform maps
the four kinematic features to approximately zero-mean, unit-variance Gaussians,
which improves training stability and the final loss, while the identification
code and vertexing features pass through unchanged.

\paragraph{Optimization.} All models are trained with AdamW
($\beta_1 = 0.9$, $\beta_2 = 0.95$)~\cite{olmo2025olmo3} and weight decay
$0.1$, with no weight decay applied to the input embeddings, normalization
gains, or any one-dimensional parameter. The learning rate follows a linear
warmup and a cosine decay to a floor of ten percent of the peak value.
Gradients are not clipped. Training runs in BF16 mixed precision with FP32
master weights, with the $z$-loss of Sec.~\ref{sec:methods-loss}.

\paragraph{IsoFLOP sweep.} The sweep trains the ladder of
Sec.~\ref{sec:methods-arch} at global batch size 512 across compute budgets
from $3\times 10^{15}$ to $3\times 10^{18}$ FLOPs, with peak learning rates
swept over $\{10^{-3}, 5\times 10^{-4}, 10^{-4}\}$ and up to three seeds per
configuration; for each (budget, model) cell the run with the lowest smoothed
loss is kept, giving the 136 fit points of Sec.~\ref{sec:extrapolation}.
Warmup is the smaller of 2000 steps and ten percent of the total steps. Runs
are constrained to between $10^{7}$ and $10^{9}$ training jets, so no sweep
run exceeds one pass over the corpus. Each sweep run trains on a single
NVIDIA A100 GPU of the Perlmutter supercomputer.

\paragraph{Large runs.} The five larger models used to test extrapolation
($15.9$M to $397$M parameters) are trained at global batch size 16384 with
peak learning rate $5\times 10^{-4}$, 2000 warmup steps, and approximately
$1.8\times 10^{5}$ steps, corresponding to $3.0\times 10^{9}$ jets or roughly
$2.8$ passes over the corpus. These runs use up to 16 Perlmutter nodes with
four A100 GPUs each under PyTorch distributed data parallelism.

\paragraph{Fine-tuning protocol.} Downstream evaluation uses the Top Quark
Tagging Reference Dataset~\cite{kasieczka2019landscape} ($1.2$M training,
$400$k validation, and $400$k test jets, providing only the four kinematic
features) and the Pythia8 quark/gluon
dataset~\cite{komiske2019energyflow} ($2$M jets balanced between quark and
gluon, split $1.6$M$/200$k$/200$k into train, validation, and test). Both are
converted to the pretraining feature layout and pass through the same frozen
Yeo-Johnson input transform. For each pretrained checkpoint we discard the
210-way head, attach a freshly initialized linear binary head, and fine-tune
the full network end to end with the optimizer settings above and separate
learning rates for the head and the pretrained trunk. We explored a grid over head learning rates
$\{10^{-4}, 10^{-5}, 10^{-6}\}$, trunk learning rates
$\{10^{-5}, 10^{-6}\}$, and batch sizes 256 and 512. The ladder-wide transfer
study of Sec.~\ref{sec:transfer} fine-tunes every selected pretrained
checkpoint at batch size 512 with head and trunk learning rates of $10^{-5}$
for $40$k steps with 2000 warmup steps, repeated five times as independent
runs. Test-set metrics are evaluated every 1000 steps. Training always runs for
the fixed 40k-step schedule, and test metrics do not alter optimizer updates,
early stopping, or training duration. For each reported quantity, however, we
average over the five runs the best test value each reaches over training
(lowest loss or highest rejection), the same protocol used for
Tables~\ref{tab:benchmarks_top} and~\ref{tab:benchmarks_qg}. This best-over-training
summary uses the test trajectory and is therefore not a validation-only model
selection protocol; it should be read as a limitation of the reported absolute
benchmark values. The ordering across the scaling ladder is evaluated under the
same protocol for every checkpoint.

\subsection{Data, code, and model availability}
\label{sec:availability}

We release all five pretrained \particlevit{} models spanning the ladder, named
\particlevit{}-S through \particlevit{}-XL (Table~\ref{tab:checkpoints}), so that
groups with different compute budgets can choose an operating point on the
scaling law. Each release includes the checkpoint, the exact preprocessing and
input-normalization constants, the pretraining configuration, the fine-tuning
protocol used for Tables~\ref{tab:benchmarks_top} and~\ref{tab:benchmarks_qg} (trunk and head learning rates,
schedule, and batch size), and a model card stating the corpus and recipe.
The models are available on the Hugging Face Hub under
\url{https://huggingface.co/collections/jaluus/particlevit} (\particlevit{}-S through
\particlevit{}-XL), and the full training, scaling-law analysis, and
fine-tuning code at \url{https://github.com/Jaluus/ParticleViT}; both are
released under the MIT license. The pretraining corpus is the publicly
available \omnilearned{} bundle~\cite{bhimji2025omnilearned}, and the two
downstream benchmarks, the Top Quark Tagging Reference
Dataset~\cite{kasieczka2019landscape} and the Pythia8 quark/gluon
dataset~\cite{komiske2019energyflow}, are public.

\begin{table}[t]
      \centering
      \caption{The five released \particlevit{} checkpoints, one per rung of the
            scaling ladder: small (S), medium (M), base (B), large (L), and extra large
            (XL). $L$ is the number of transformer blocks and $d$ the model width; all
            sizes use a head dimension of $64$ (so the head count is $d/64$), SwiGLU
            feedforward layers of width $d_{ff} = \lfloor \tfrac{8}{3}\,d \rfloor$, a single
            class token, no positional encoding, and a 210-class pretraining head. $N$ is
            the total trainable parameter count and $C$ the pretraining compute, accounted
            as six FLOPs per parameter per token over $3.0\times10^{9}$ jets at the measured
            mean occupancy of the 150 particle slots. These checkpoints are the five models
            of Tables~\ref{tab:benchmarks_top} and~\ref{tab:benchmarks_qg}, there rounded to $16$--$397$M.}
      \label{tab:checkpoints}
      \begin{ruledtabular}
            \begin{tabular}{l ccc r l}
                  Name              & $L$ & $d$  & heads & $N$      & $C$~[FLOPs]        \\
                  \colrule
                  \particlevit{}-S  & 5   & 512  & 8     & 15.9\,M  & $1.4\times10^{19}$ \\
                  \particlevit{}-M  & 7   & 640  & 10    & 34.6\,M  & $3.0\times10^{19}$ \\
                  \particlevit{}-B  & 8   & 832  & 13    & 66.7\,M  & $5.7\times10^{19}$ \\
                  \particlevit{}-L  & 10  & 1024 & 16    & 126.1\,M & $1.1\times10^{20}$ \\
                  \particlevit{}-XL & 14  & 1536 & 24    & 396.8\,M & $3.4\times10^{20}$ \\
            \end{tabular}
      \end{ruledtabular}
\end{table}

\section*{Note added}

As this manuscript was being finalized, we became aware of a recent ATLAS note~\cite{atlas2026carpedatum} which also derives joint model-data scaling laws for jet tagging.
That note tests their predicted performance gains on heavy-flavor tagging with an 86-million-parameter classifier trained on 7.7 billion jets.
Our approach is similar, but we consider multiple downstream tasks and correlate pretraining loss with downstream performance.
We have also made all of our code and data publicly available.\\

\begin{acknowledgments}
We thank Divya Nori, Michael Zhang, Tanvi Wamorkar, Gregor Kržmanc,
Alkaid Cheng, and Kevin Greif for useful discussions and feedback.
This work was supported in part by the U.S. Department of Energy (DOE)
under Contract No.~DE-AC02-76SF00515. This research used resources of the
National Energy Research Scientific Computing Center (NERSC), a DOE Office
of Science User Facility supported by the Office of Science of the U.S.
Department of Energy under Contract No.~DE-AC02-05CH11231, under NERSC
award HEP-ERCAP0035546.

\textit{AI usage statement.} OpenAI Codex (GPT-5.5, GPT-5.6) and Claude Code (Opus 4.8, Fable) were used to assist
with software development, run monitoring and manuscript preparation. The authors directed
the agents and manually reviewed and verified their outputs. The authors
take full responsibility for the accuracy and integrity of this work.
\end{acknowledgments}

\bibliographystyle{apsrev4-2}
\bibliography{refs}

\end{document}